\documentclass[conference]{IEEEtran}
\IEEEoverridecommandlockouts
\usepackage{graphicx} 
\usepackage{multicol}
\usepackage{amsmath}
\usepackage{array}
\usepackage{cite}
\usepackage{amsmath,amssymb,amsfonts}
\usepackage{amsthm}
\usepackage{wrapfig}
\usepackage{algorithm}
\usepackage{algpseudocode}

\usepackage{graphicx}
\usepackage{textcomp}
\usepackage{xcolor}
\usepackage{multirow}
\usepackage{booktabs}
\usepackage{soul}
\usepackage{array}
\usepackage{float}
\usepackage{tikz}
\usepackage{longtable}
\usepackage{fancyhdr}
\usepackage{nicematrix}
\usepackage{graphicx}

\title{Cross-temporal Detection of Novel Ransomware Campaigns: A Multi-Modal Alert Approach}
\author{Sathvik Murli, Dhruv Nandakumar, Prabhat Kumar Kushwaha, Cheng Wang, \\Christopher Redino*, Abdul Rahman, Shalini Israni, Tarun Singh, Edward Bowen \\
Deloitte \& Touche LLP  \\ * Corresponding Author: credino@deloitte.com
}
\date{May 2023}

\begin{document}

\maketitle

\begin{abstract}
We present a novel approach to identify ransomware campaigns derived from attack timelines representations within victim networks. Malicious activity profiles developed from multiple alert sources support the construction of alert graphs. This approach enables an effective and scalable representation of the attack timelines where individual nodes represent malicious activity detections with connections describing the potential attack paths. This work demonstrates adaptability to different attack patterns through implementing a novel method for parsing and classifying alert graphs while maintaining efficacy despite potentially low-dimension node features.
\end{abstract}


\section{Introduction}
Ransomware is a form of attack in which an individual or group of attackers gain access to a network or devices with sensitive data on it. They then steal this data and encrypt it, demanding a ransom for its safe return. Ransomware attacks have also evolved over time, moving from just encrypting data to extorting victims by also exfiltrating data. As explored in \cite{razaulla2023age,oz2022survey}, ransomware attacks commonly occur in several stages. These stages generally involve the initial disbursement of malware, infection of a target network, staging of ransomware, and finally encryption of sensitive data. Encryption of sensitive data can take many forms. For example, in a Doxware attack \cite{razaulla2023age}, hackers may threaten to make user’s private data public. This causes irreversible damage as this data can never be made private again. Or in a traditional crypto-ransomware attack, data will be encrypted without necessarily affecting user’s systems.  \par

This general structure can still allow for different types of attacks and intrusions to occur during a ransomware campaign. Ransomware taxonomies often show multiple attack types within the infection stage for example, and the same can be said of command and control (C\&C) communication\cite{moussaileb2021survey,razaulla2023age}. This applies to all stages of ransomware attacks, with cases being observed where Distributed Denial of Service (DDoS) attacks are used to disguise data exfiltration, or banking trojans become precursors to malware payloads. Given these different attack patterns, ransomware attacks become more difficult to detect ahead of the final encryption phase with piecemeal detections of different attacks.
  In this work, we hypothesize that a modelling approach that considers disparate alerts over a large time frame as part of a larger campaign would be able to effectively detect ransomware attack at various stages of the campaign.  \par


In this paper we present a novel approach to identify ransomware campaigns based off  representations of attack timelines as they occur on victim networks. The key contributions of this paper are two-fold:
\begin{itemize}
\item An effective and scalable representation of the attack timelines, formed by combining malicious activity detections coming from multiple alert sources into alert graphs. Within these graphs, individual nodes represent malicious activity detections and the connections describe the potential attack paths
\item A method for parsing and classifying alert graphs, that maintains efficacy despite potentially low-dimension node features
\end{itemize}
Our approach will allow us to detect ransomware attacks throughout their campaigns, and adapt to different potential attack paths and patterns. \par

\section{Review of the Literature}
Given the ubiquity of these attacks and the speed at which they occur, a wide range of solutions have been developed. As discussed in \cite{razaulla2023age}, these solutions vary based on the data they are analyzing and the analysis or detection method. Several research methods have been proposed using static and dynamic analysis, both of which involve analyzing features like opcodes, hashes, and byte sequences of ransomware files. Static analysis inspects a suspicious binary without executing it, while dynamic analysis involves investigation of a binary as it is executing in a “sandbox” environment \cite{razaulla2023age}. \par

Many rule-based approaches to ransomware detection have been proposed.  Most of these methods use a variety of input features to detect ransomware file execution. 

Scaife et al. \cite{scaife2016cryptolock} and Naik et al. \cite{naik2020fuzzy} both developed rulesets by operating on indicators based on files and filesystem behaviors. Morato et al. \cite{morato2018ransomware} presented an algorithm, Ransomware Early Detection from File SHaring Traffic (REDFISH), based on analysis of network traffic to develop rules that would indicate potential encryption of files in shared network volumes. Others have focused more on decoy-based systems such as honeypots, like  Moore \cite{moore2016detecting} and G{\'o}mez-Hern{\'a}ndez el al.\cite{gomez2022inhibiting}. 

Multiple machine learning or deep learning based methods have been proposed to combat ransomware attacks as well, largely focusing on static or dynamic analysis. 

Poudyal et al. \cite{poudyal2019multi}, Zahoora el al. \cite{zahoora2022zero}, Sgandurra el al. \cite{sgandurra2016automated} all proposed methods operating on function and application programming interface (API) calls, leveraging methods such as N-gram generation, a deep contractive autoencoder, and regularized logistic regression respectively. Zhang et al. \cite{zhang2020ransomware} developed a Self-Attention based Convolutional Neural Network to create intermediate features based off N-gram embeddings of disassembled opcodes, which are used to classify ransomware files. The aforementioned methods rely on static or dynamic analysis, which can neglect other stages in ransomware campaigns that may present attackers with opportunities to evade detection systems that operate on binaries.\par
 
Our proposed method for detecting ransomware campaigns using representations of attack timelines as alert graphs allows us to identify the goals of attackers based off the potential attack patterns they are using. Wang et al. \cite{wang2022threatrace} employed a similar method to identify the host-based threats by constructing provenance graphs from function calls and filesystem operations. They then employed inductive representation learning, GraphSAGE, to classify actions as malicious or benign. However, their methodology focused on tracing actions on individual hosts based on function calls whereas  we consider a holistic representation of entire networks, in order to track ransomware attacks potentially targeting and infecting multiple hosts, across their timeline, no matter what stage they are currently executing. \par



\section{Data Structuring}

\subsection{Simulation}As has been identified by \cite{razaulla2023age , oz2022survey}, a majority of previous research efforts have relied on datasets from the encryption stage such as the VirusShare dataset \cite{VirusShare} or the VirusTotal \cite{VirusTotal} dataset. 
To track ransomware across its entire timeline, the dataset we use must capture data across a ransomware campaign. During these campaigns several malicious intrusion methods may raise alerts in different cybersecurity tools, and our approach correlates and combines these alerts to form representations of attack campaigns, that we classify as ransomware. \par 

To create our dataset, we leverage an industry standard attack simulation tool to simulate ransomware campaigns on a set of hosts which have industry standard endpoint detection and rules-based network detection sources. The hosts will also run network flow data through a malware severity ranking (MSR) model \cite{nandakumar2023foundational} which has been shown in prior literature to identify anomalous network flow characteristics and malware executions with strong performance. To form the training and testing set, we simulate several ransomware campaigns, and a set of other types of malware campaigns. 
These campaigns are listed in Table \ref{tab:RWareCampaignList}. During the execution of these simulations, data is collected from the listed alert sources and combined into a unified representation to form our input dataset. This is fed to our algorithm for constructing a representation of attack campaigns as alert graphs.\par

\begin{table*}
\centering
\caption{Campaign List used to construct Training Set}
\label{tab:RWareCampaignList}
\vspace{1.5ex}
\scalebox{.85}{
    \begin{tabular}{|p{2cm}|p{3cm}|l|p{1.5cm}||p{2cm}|p{3cm}|l|p{1.5cm}|}\hline\hline
       {\bf Campaign Name}  &  {\bf Campaign Inspiration} & {\bf Category} & {\bf Number of Constructed Graphs} & {\bf Campaign Name}  &  {\bf Campaign Inspiration} & {\bf Category} & {\bf Number of Constructed Graphs} \\\hline
       Yanluowang & Yanluowang ransomware gang’s attack  & Ransomware & 839 & APT 29 & Behaviors of APT 29 threat group & Other Malware & 1451\\\hline 
       Wannacry & Wannacry cryptoworm & Ransomware & 262 & APT 39 & Behaviors of APT 39 threat group & Other Malware & 1451\\\hline
       Maze Ransomware & Infection and propagation of Maze & Ransomware & 7759 & FireEye & Against FireEye assessment tool & Other Malware & 1133\\\hline
       MuddyWater & Attack against Tecnion Israel Institute of Technology & Ransomware & 777 & AppleJeus & US-CERT Alert AA21-048A & Other Malware & 1292\\\hline
       K12 & US-CERT Alert AA20-345A & Ransomware & 1014 & Badcall & Campaign used to turn machines into relay points & Other Malware & 194\\\hline
       DeepBlueMagic & IL-CERT Alert 1389  & Ransomware & 198 & Pulse Secure & US-CERT Alert AA20-107A & Other Malware & 95\\\hline
       Cuba & Cuba ransomware  & Ransomware & 464 &  Hidden Cobra & US-CERT Alert AA20-106A & Other Malware & 589\\\hline
       Conti & US-CERT Alert AA21-265A  & Ransomware & 529 &  Cert AA22-216A & US-CERT Alert AA22-216A, multiple different malware strains & Other Malware & 1634\\\hline
       Clop & Campaign targeting windows machines & Ransomware & 712 &  Qsnatch & Qsnatch malware & Other Malware & 245\\\hline
    \end{tabular}

}
\end{table*}

\subsection{Unified Representation} In developing this methodology, we intend
 to remain source and tool agnostic. In the cybersecurity industry, there are a plethora of tools that enterprises may use to help protect their networks, which will generate alerts and information that the proposed approach could leverage. In order to ensure our algorithm can support and integrate these alerts with varying formats and characteristics, we intend to construct a unified representation that will support different types of alerts and detections. For each attack campaign, we observe the data from each of the sources mentioned in the previous section have a timestamp, risk score which quantifies the impact of a certain event relative to the network it originates in, category which indicates what type of malware violation it is, and a severity score which quantifies the event's severity across an industry standard baseline such as the Common Vulnerability Scoring System (CVSS). These pieces of information become the node features that we use to represent each individual alert. Each alert can also be described by a source host and, in some cases, a destination host. 
 Only if an alert involves a connection of some kind to another host, or some form of authentication event, then the other host becomes the destination associated with the alert. We generate alert graphs for each individual source host, and build connections between these alert graphs based on the associated destination hosts. Once connections are built between individual sources' host graphs to form a larger graph, this becomes the input to our model. 
Our approach is highly extensible as we can incorporate various alert sources from other tools in addition to the ones used in this study, since we are not using Internet Protocol (IP) addresses or any features which are dependent upon a particular tool or sensor.\par


\subsection{Alert Graph Construction} 
\begin{figure*}
  \center
  \includegraphics[scale = .375]{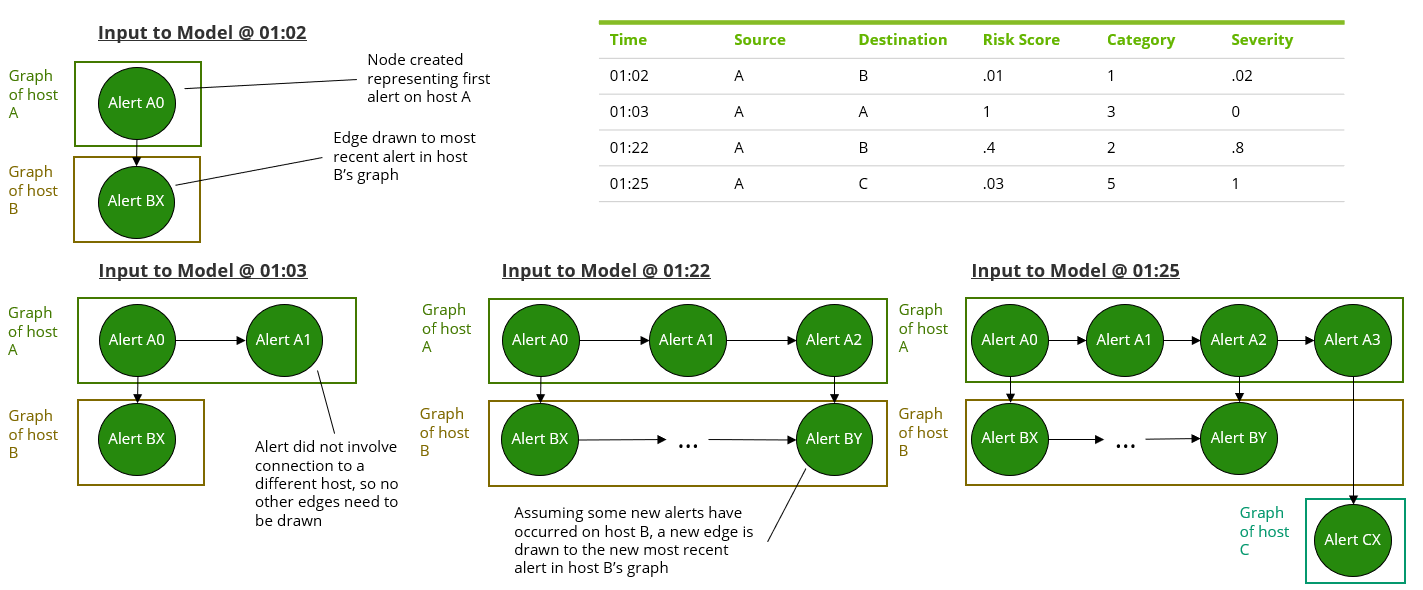}
  \caption{Algorithm used to construct input graphs. Shown here is an example set of alerts to add to existing alert graphs, and the progression of the graphs associated with these alerts. The label on each node refers to the host and the order in which the alert arrived. For example, A0 is the first node in the graph of host A. BX is the most recent node in the graph of host B.}
  \label{fig:graphalgo}
\end{figure*}

Shown in Fig. \ref{fig:graphalgo} is our process for constructing alert graphs, from an example table of detections. The example alerts in the figure have already been transformed to follow our unified representation. The timestamp, risk score, severity score, and category will become our node features, whereas the source host and destination host are used to determine what directed edges are drawn. 
The alert sources are normalized using MinMaxScaling, such that each alert is normalized with respect to the minimum and maximum values of detections observed from each tool. Our algorithm considers alerts in batches. Each time an alert is received we update our graph representation of the overall campaign, and this new representation, with the newly added alert as a node and associated edges, becomes the input to the model. This gives us a distinct graph for each timestep and alert we intend to analyze. \par 

Each host receives its own alert graph, and before being input into our model these graphs are combined based off the connections that are part of each alert. This process is shown by the example input set in Figure \ref{fig:graphalgo}, where each row has an associated update to the existing alert graphs.

As the attacks we analyze move to their next stages, more alerts are received and are again joined to the previously constructed graphs. Our alert graphs grow as the number of alerts and detections increase, increasing the nodes and edges drawn and the number of timesteps included. So as the graph size increases, the detection range and timeline of the attack also increases to its next stage. 
It is important to note, here, that our approach does not require the tools from which alerts originate to be tuned to ransomware detection specifically. In fact, the three data sources listed above and used in this work are not tuned to such tasks and produce general alerts for most kinds of suspicious activity including false positive events. However our modelling approach shows strong performance despite these potential false positive events, which shows our ability to assist with potential weaknesses of the underlying tools used as inputs. It also highlights the effectiveness of our methodology to potentially noisy data sources, allowing for easier deployment to existing enterprise environments.\par

\section{ Methodology }


\subsection{Pipeline}
\subsubsection{Initial Embedding Architecture}
To create graph embeddings that can be used as inputs into our classifier, we implement the architecture described in Fig \ref{fig:modarch}.  We leverage graph isomorphism (GIN) layers as described in \cite{powerful}. 


Our graph isomorphism layers use a single layer multi-layer perceptron (MLP) to update node representations between layers. These representations are then fed into our choice of READOUT function, a positional encoding layer and a transformer encoding layer as described in \cite{vaswani2017attention}, to create a graph-level representation. This transformer encoding layer operates on the node representations generated for the final 120 nodes of each of our alert graphs. In the context of our representation of attack campaigns, this means the transformer encoding layer is operating on the most recent 120 alerts. \par

\subsubsection{Subsequent Embedding Architecture} For the second task that is part of our training algorithm, we include 2 sets of 2-dimensional (2D) convolutional layers and 2D max pooling layers with our initial embedding architecture. These convolutional layers operate on the outputs of the initial embedding architecture. 3 versions of the previously described GIN layers and attention-based pooling will be concatenated to form the inputs into the convolutional layers.
We use the outputs of 3 versions of GIN layers together, treating them like one "image" of the graph's properties. One set of convolutional layers is intended as a projection head to be trained on campaign classification, and the other acts as a projection head to be trained on graph contrastive learning (GraphCL), as described in \cite{you2020graph}. These two training tasks occur in sync, with 2 cross-stitch units placed between them as described in \cite{misra2016cross}. These units 
share information between the 
projection heads, to increase their respective performance levels by identifying useful pieces of information from the other stacks. \par

\subsubsection{Classifier Architecture}
As the projection heads are being trained, the value of the information computed by the GIN layers in the initial embeddings increases. As a result, we can utilize their outputs to perform our graph classification task. The GIN layers are fed into a new choice of READOUT function, a summation aggregator that adds all the node representations to form a graph representation which becomes the input into a set of linear layers and rectified linear unit (ReLU) activations that output a decision on whether the campaign being examined is a ransomware campaign. \par

\subsection{Sampling Algorithm}

Our campaign simulations result in a total of 20638 graphs, as shown in \ref{tab:RWareCampaignList}. Our graph construction algorithm creates alert graphs each campaign. We evaluate our model architecture by constructing our training set using two methods. We create a training set consisting only of specific campaigns from our overall set, and a testing set consisting of the remaining campaigns. 
We also evaluate our model's performance specifically against the first 20\%, 40\%, and 60\% of graphs included in each campaign in the testing set, which correspond to the first 20\%, 40\%, and 60\% of a campaigns timeline. 
We found through experimentation that we had the strongest performance consistently on training sets that were comprised of graphs corresponding to ransomware and malware respectively in a 1:1 ratio, or as close as possible. \par



\begin{figure*}
    \begin{center}
        \includegraphics[width=\textwidth]{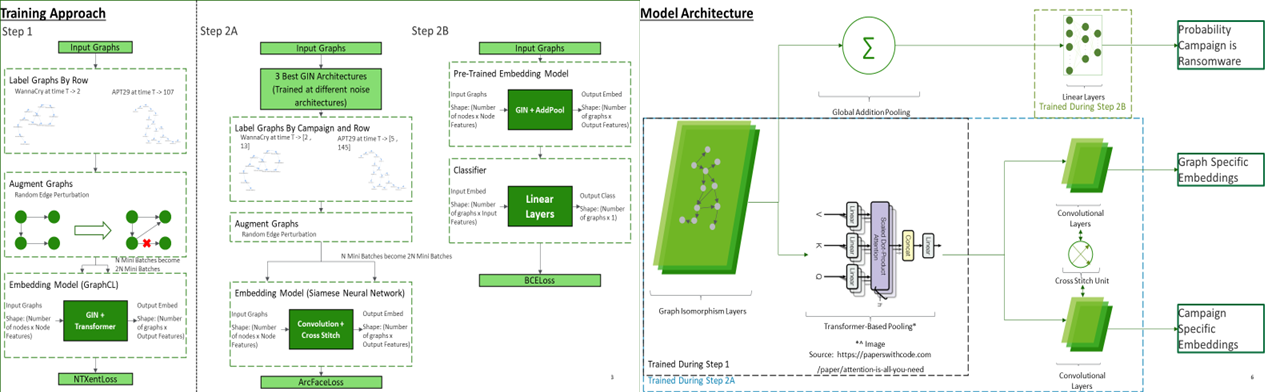}
        \centering
        \caption{Model Architecture with Associated Training Steps. (Left) Shown here is our training algorithm. (Right) Shown here is our model architecture along with the training steps that affect each layer.}
        \label{fig:modarch}
    \end{center}
\end{figure*}


\subsection{Training Algorithm}

\subsubsection{Initial Embedding Training}
Our general architecture along with which training steps apply to which layers is shown in Figure \ref{fig:modarch}.
The first step in training our embedding architecture is labelling each graph individually, so each graph has a distinct identifier. Our algorithm for training the embedding architecture involves two forward propagations through the same network. For each of the two forward passes, noise is added using edge perturbation to the data in the input batch. A percentage of the edges are removed from the input graphs, and a smaller percentage of random edges are added to the graphs. Then the embeddings are passed into the normalized temperature-scaled cross entropy loss (NT-Xent) \cite{chen2020simple}. The addition of noise in the form of edge perturbation allows our model to generalize to a wider array of input graphs, and strengthens our performance on previously unseen campaigns. This particular step is referred to as step 1 in Fig. \ref{fig:modarch}.  

\subsubsection{Subsequent Embedding Training and Classification Training} The next steps in our training process occurs simultaneously. First the model is trained as a Siamese neural network, where the inputs to the models are graphs that have been labelled by which campaign simulation they refer to. Edge perturbations are again applied to the graphs before they are fed into the head dedicated to that task, and the loss of these forward passes is computed using the Sub-center ArcFace Loss \cite{deng2020sub} for an entire epoch. Then the initial training step is repeated but with 
 the forward pass going through the projection head this time. This step is referred to as step 2A in Fig. \ref{fig:modarch}. While these two computations are occurring, the cross-stitch units learn what information should be shared between the tasks. After these metric learning tasks, an epoch is completed for the classification task. But as described earlier, 
the linear layers are trained to learn to classify the graphs based off of the outputs of the frozen GIN layers. This classification step is step 2B in Fig. \ref{fig:modarch}. 

\subsubsection{HyperParameter Tuning}
As part of our algorithm, there are several hyper-parameters available to tune. In general, for each step in our training algorithm we use the Nesterov-accelerated Adaptive Moment Estimation (NAdam) optimization algorithm. The first is the graph contrastive learning optimizer used in Step 1, and re-initialized for both projection heads in Step 2A. We found the strongest, consistent performance using a learning rate of .00085 and a weight decay of .000085 for both of these optimizers. 
The Sub-center ArcFace Loss function, also requires its own NAdam optimizer with the same parameters. For the classification step, Step 2B, 
the NAdam optimizer had varying learning rates. In training runs where the training set was comprised of multiple smaller campaigns, we found a smaller learning rate and weight decay combination was most effective, .00045 and .000025 respectively.
For training sets with larger campaigns, a combination of .0025 and .000025 for learning rate and weight decay was used. \par

\subsubsection{Motivation}
The purpose of this multiple step training algorithm is to create a model that can generalize well to ransomware or malware campaigns it may have not yet seen, while maintaining an understanding of the campaigns it has already learned to identify as ransomware. Once our alert graphs have been constructed, we obtain an input set comprised of graphs with low dimension node features, and highly variable structures. 
Due to the low dimensionality of our node features extracting valuable information in the form of node embeddings becomes difficult. Our initial graph contrastive learning approach, Step 1, helps increase the fidelity of our node embeddings. However, we also need to ensure that our graph embeddings are also extracting information that is valuable to identifying campaign specific information, to classify these graphs as part of ransomware campaigns or not. This necessitates our second training step, Step 2A, where we cross-stitch the Siamese neural network with the original graph contrastive learning task. We train our classifier in sync with the Siamese neural network task and contrastive learning task, on frozen versions of our graph embeddings. Through experimentation, we found that allowing the GIN layers 
to learn from the classification task as well resulted in weaker performance on all the tasks that are part of our training algorithm. We also train our classifier at the same time as the contrastive learning and Siamese neural network tasks in step 2B because we found that this training approach prevented severe overfitting to any one version of the graph embeddings, which would severely limit our abilities to classify previously unseen campaigns. \par

\section{Results}

\begin{table*}
\centering
\caption{Testing Splits \& Results}
\label{tab:TestingSplitResults}
\scalebox{1}{
    \begin{tabular}{p{7cm}|p{7cm}|p{.5cm}|p{.5cm}|p{.5cm}}\hline \hline
         {\bf Training Campaigns} & {\bf Testing Campaigns} & {\bf Prec- ision} & {\bf Re- call} & {\bf AUC Score} \\ \hline
         Conti, MuddyWater, Yanluowang, Clop, DeepBlueMagic, Cuba,  Apt29, Apt39, FireEye & Maze, K12,WannaCry, Badcall, PulseSecure, Cobra, Cert, Applejeus, Qsnatch & .94 & .93 & 0.932  \\ \hline
         Conti, MuddyWater, Yanluowang, Cuba,  DeepBlueMagic, K12, Apt29, Apt39, FireEye & Maze, Clop, WannaCry, Badcall, PulseSecure, Cobra, Cert, Applejeus, Qsnatch & .90 & .90 & 0.904  \\ \hline
         Clop, K12, Maze(first 12\% of graphs), MuddyWater, DeepBlue, Cuba, PulseSecure, BadCall, Apt29, FireEye & WannaCry, Conti, Yanluowang, Apt39, Qsnatch, Applejeus, Cert & .91 & .91 & .90\\\hline
         MuddyWater, DeepBlue, Conti, K12, Yanluowang, Cuba, PulseSecure, Badcall, Cert, Cobra, AppleJeus, Qsnatch & Maze, WannaCry, Apt29, Apt39, FireEye & .84 & .94 & .89\\\hline
    \end{tabular}
}
\end{table*}

\begin{figure*}
    \begin{center}
        \includegraphics[width=\textwidth]{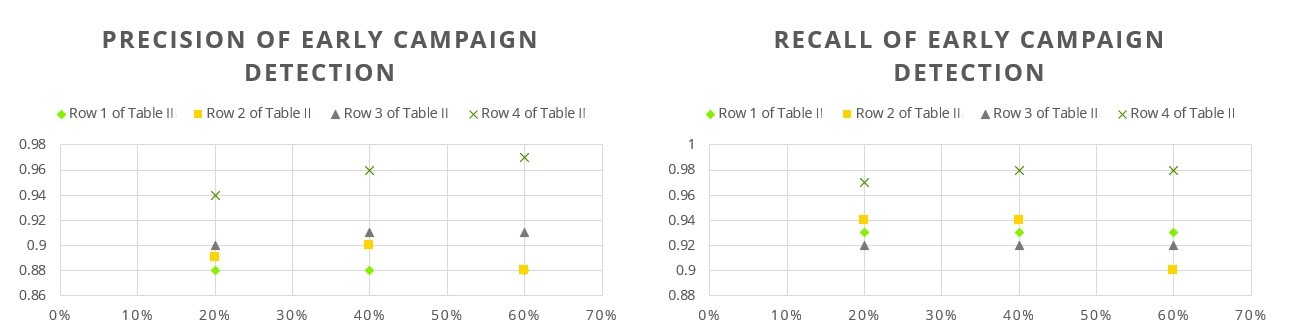}
        \centering
        \caption{Early Campaign Detection Statistics. Shown in these plots are statistics for each of our chosen testing splits for the first 20\% 40\% and 60\% of all the campaigns in the testing set.}
        \label{fig:earlycampaign}
    \end{center}
\end{figure*}

Our modelling approach has shown strong performance on both versions of the testing sets we consider. In order to evaluate the performance of our model, we focused on 3 key metrics: precision, recall, and area under the curve (AUC) score. 
Examples of these results are shown in Table \ref{tab:TestingSplitResults}.
From our experiments, we see that our model has an average AUC score of .9 for experiments that consist of different combinations of unseen ransomware and malware campaigns. These results show the strength of our modelling approach on identifying newer ransomware attack vectors that malicious actors may employ because our model can perform well on identifying unseen campaigns. 
For each new training set that was used as part of our experiments, minimal tuning of the ransomware classification learning rate was required to achieve strong performance. 
\par 

Of particular interest in our experimentation is detection of earlier stages of ransomware campaigns. Most detection methods are centered around detection of ransomware executables or ransomware encryption. To determine the potential of our approach, we test our model against the first 20\%, 40\%, and 60\% of graphs associated with each campaign used in our testing sets. Based on our graph construction algorithm, these graphs would correspond to the earlier stages of ransomware or malware campaigns. Shown in Fig. \ref{fig:earlycampaign} are results of evaluating our model against the aforementioned ratios of the testing campaigns. Our performance holds for earlier stages of ransomware and malware campaigns, demonstrating the efficacy of our approach in assisting with early prevention of ransomware campaigns. \par

There are certain caveats when investigating our performance. We can identify unseen campaigns when our training set has a campaign of similar or larger size in terms of number of alerts or stages. When we train our model on smaller campaigns exclusively, performance falters when evaluating against larger ransomware or malware campaigns. However, when trained on a mix of large and small campaigns, our model retains its strong performance.  \par

\subsection{Baseline Modeling}

\begin{table*}
\centering
\caption{Base Classification Approaches \& Results}
\label{tab:ClassifierResults}
\vspace{1.5ex}
\scalebox{1.1}{
    \begin{tabular}{c|c|c|c}\hline \hline
         {\bf Approach} & {\bf Precision} & {\bf Recall} & {\bf AUC Score} \\ \hline
         Adjusted GCN & .55 & .50 & .5005  \\ \hline
         VGAE & .385 & .50 & .5  \\\hline
         GIN & .55 & .50 & .5005 \\\hline
    \end{tabular}
}

\centering
\caption{Metric Learning Based Approaches \& Results}
\label{tab:MetricResults}
\vspace{1.5ex}
\scalebox{.9}{
    \begin{tabular}{l|l|l|l}\hline \hline
         {\bf Approach} & {\bf Added Noise} & {\bf Strongest AUC Score} & {\bf Embedding Task} \\ \hline
         Adjusted GCN & Edge Perturbation (15\% to 35\%) & .85 & Campaign  \\ \hline
         Adjusted GCN & Node Dropping (15\% to 35\%) & .74 & Campaign \\ \hline
         Adjusted GCN & Node Dropping (15\% to 35\%) & .55 & Graph  \\ \hline
         Cross-Stitched GCN's & Edge Perturbation (20\% to 45\%) & .65 & Campaign/Graph \\\hline
         GIN & Edge Perturbation (15\% to 35\%) & .7 & Campaign \\\hline
         GIN with Transformer based Pooling & Edge Perturbation (15\% to 35\%) & .7 & Campaign \\\hline
    \end{tabular}
}
\end{table*}

Shown in Table \ref{tab:ClassifierResults} and Table \ref{tab:MetricResults} are some of the other methods tested for the ransomware classification problem. As shown by our results, other methods' performance suffer in comparison. 

The first two rows of results shown in Table \ref{tab:ClassifierResults} denote performance using a more standard classification approach, with BCELoss. Our baseline approaches' performance suffered severely on testing sets 
built from unseen campaigns. 
The first row of results in particular denotes a graph convolutional network (GCN) where the message construction algorithm was as described in \cite{wang2019dynamic}. The second row of results shows an approach where a VGAE or variational graph autoencoder as described in \cite{kipf2016variational}. This approach was constructed using MSELoss or Mean Squared Error Loss to compute the models ability to reconstruct its inputs, with the hypothesis that ransomware graphs would be constructed more effectively than malware graphs. \par 

The results shown in Table \ref{tab:MetricResults} denote training runs that followed similar metric learning approaches to our current approach. The procedure followed was similar to the procedure used in \cite{you2020graph}, utilizing NT-Xent Loss on the task of constructing embeddings. The models were each trained on the  embedding task specified.
Then linear classifiers were trained to identify ransomware campaigns based on the embeddings that had been constructed. Also shown in the table are the types of noise used.  As shown by our results, while there were some experiments that yielded strong performance, the majority suffered when faced with campaigns that were previously unseen by the model. \par

\section{Conclusion, Limitations, Future Work}

In this work, we have introduced a multi-modal approach to detection of ransomware campaigns. We introduced a method of combining alerts from different detection sources into a unified, scalable representation of attack campaigns, lowering the effort needed to integrate different detection sources. 
This approach lends itself to tool agnostic implementation, elevating existing detection capabilities, enabling efficient triage and incident response, and reducing operator fatigue.  We also use graph contrastive learning-based methods to classify our alert graph representations as indicative of ransomware attacks or not. Our modelling approach shows strong performance on earlier stages of ransomware campaigns, where some previously investigated detection methods have faltered. We also have strong performance on campaigns that are previously unseen to our model, which is crucial to an approach of this kind. 
In order to remain effective against these a constantly evolving cyber-threat landscape, a modelling approach must be able to identify similarities between threats whether or not it has been trained against them, and our approach has that capability.\par

We also believe this work can be expanded up on in the future. Our usage of alert graphs allows us to potentially identify other information about attack campaigns, such as  
identifying what stage of an attack is taking place. We can use our graph representations to start predicting future alerts that may be raised as part of attack campaigns. This could help operators and threat hunters proactively identify vulnerabilities. Our strong results thus far show that this approach can be a foundation for multiple tools and models that will assist in combatting different cyber threats in the future.Furthermore, we believe that the model architectures and training methodologies proposed in this work are also more broadly applicable to graph-based problem spaces wherein the population of graphs are hierarchically organized by classes and sub-classes; similar to how our problem was organized by ransomware/malware and their constituent campaigns.  \par








\bibliographystyle{IEEEtran}
\bibliography{ref}
\end{document}